\documentclass[a4paper,12pt]{article}
\usepackage[top=1.25in, bottom=1.25in, left=1.05in, right=1.05in]{geometry}
\usepackage[utf8]{inputenc}
\usepackage[T1]{fontenc}
\usepackage[english]{babel}

\usepackage{natbib}
\usepackage[section]{placeins}
\usepackage{setspace}

\usepackage{enumitem}
\setlist{noitemsep}  

\usepackage{amscd,amsmath,amssymb,amsfonts,amsthm,bbm,bm,latexsym,mathrsfs,mathtools,bigints}
\usepackage[subnum]{cases}	
\usepackage{dsfont}
\makeatletter
\newcommand*{\distas}[1]{\mathbin{\overset{#1}{\kern\z@\sim}}}	
\makeatother

\allowdisplaybreaks

\theoremstyle{remark}

\theoremstyle{plain}

\RequirePackage[colorlinks=true, citecolor=red, urlcolor=blue]{hyperref}

\usepackage[dvipsnames]{xcolor}
\usepackage{subfigure,epsfig,epstopdf,rotating,graphics,graphicx} 
\graphicspath{{./figures/}}
\usepackage[width=0.9\textwidth, font={footnotesize}]{caption}

\usepackage{rotating,lscape,pdflscape}
\usepackage{booktabs,longtable,float,array,multirow,colortbl,adjustbox}
\usepackage{threeparttable}

\usepackage[colorinlistoftodos,color=orange!60]{todonotes}
\usepackage{regexpatch}
\makeatletter
\xpatchcmd{\@todo}{\setkeys{todonotes}{#1}}{\setkeys{todonotes}{inline,#1}}{}{}
\makeatother

\makeatletter

\def \bu{\mathbf{u}}
\def \bw{\mathbf{w}}
\def \bx{\mathbf{x}}
\def \by{\mathbf{y}}
\def \bz{\mathbf{z}}

\def \bH{\mathbf{H}}

\def \bX{\mathbf{X}}
\def \bY{\mathbf{Y}}

\def \bbeta    {\boldsymbol{\beta}}
\def \bgamma   {\boldsymbol{\gamma}}

\def \boldeta  {\boldsymbol{\eta}}

\def \bmu      {\boldsymbol{\mu}}

\def \bphi     {\boldsymbol{\phi}}

\def \Id {\mathbf{I}}
\def \I {\mathbb{I}}
\def \R {\mathds{R}}

\makeatother

\title{\vspace{-60pt} \textbf{A Quantile Nelson-Siegel model}
\thanks{The authors would like to thank Gary Koop and Dimitris Korobilis for their valuable comments on an earlier version of this paper. We also thank seminar and conference participants at the University of Leicester, the Central Bank of Ireland, and the 2025 IAAE Conference for their helpful feedback and suggestions.}
}

\author{Matteo Iacopini\thanks{Luiss University, Italy. \color{blue}\texttt{miacopini@luiss.it}}\and
Aubrey Poon\thanks{University of Kent, United Kingdom and \"Orebro University, Sweden. \color{blue}\texttt{a.poon@kent.ac.uk}}\and
Luca Rossini\thanks{University of Milan, Italy and Fondazione Eni Enrico Mattei. \color{blue}\texttt{luca.rossini@unimi.it}}\and
Dan Zhu\thanks{Monash University, Australia. \color{blue}\texttt{dan.zhu@monash.edu}}
}

\date{\today}

\begin{document}

\maketitle

\begin{abstract}

We propose a novel framework for modeling the yield curve from a quantile perspective. Building on the dynamic Nelson–Siegel model of \citet{diebold2006macroeconomy}, we extend its traditional mean-based approach to a quantile regression setting, enabling the estimation of yield curve factors—level, slope, and curvature—at specific quantiles of the conditional distribution. A key advantage of our framework is its ability to characterize the entire conditional distribution of the yield curve across maturities and over time. In an empirical analysis of the U.S. term structure of interest rates, our method demonstrates superior out-of-sample forecasting performance, particularly in capturing the tails of the yield distribution—an aspect increasingly emphasized in the recent literature on distributional forecasting. In addition to its forecasting advantages, our approach reveals rich distributional features beyond the mean. In particular, we find that the dynamic changes in these distributional features differ markedly between the Great Recession and the COVID-19 pandemic period, highlighting a fundamental shift in how interest rate markets respond to distinct economic shocks.

\end{abstract}


\clearpage

\doublespacing

\section{Introduction}


The ongoing global challenges stemming from the COVID-19 pandemic and recent geopolitical conflicts have prompted a renewed focus on inflation and monetary policy. Of particular significance is the information embedded in the term structure of interest rates, offering valuable insights into the evolution of these economic factors. The slope of the yield curve has gained considerable recognition as a key predictor of economic activity, exerting a substantial influence on investors' decision-making processes.

This significance is underscored by research from \cite{Benzoni18} and \cite{Haubrich2021}, which highlights a strong correlation between an inversion of the yield curve and the onset of recessions. For instance, in May 2019, the yield curve inverted almost a year before the commencement of the most recent recession in March 2020. More recently, \cite{ECB2023} emphasised the unprecedented inversion in the risk-free yield curves in the United States due to the rapid rise in short-term interest rates since 2022. They pointed out that ``\textit{recession probability models based solely on the yield curve slope currently point to elevated odds of recession in the euro area and the United States in one year's time}'' thereby stressing the critical importance of dealing with and studying the yield curve in contemporary macroeconomic analysis.

Transitioning to a macroeconomic perspective, significant time variations have been shown to influence the dynamics of US inflation and real activity \citep{mumtaz2009time}. \cite{monch2012term} studied the evolution of macroeconomic variables by using level, slope, and curvature factors, finding evidence of the informativeness of the curvature factor about the future evolution of the yield curve. Moving to factor analysis, \cite{coroneo2016unspanned} provided evidence that some common factors drive the dynamics of macroeconomic variables and government bonds. More recently, \cite{Fernandes2019} predicted the yield curve using different forward-looking macroeconomic variables and highlighted the importance of these variables in forecasting the yield curve.


In the context of modelling the term structure of interest rates, the Nelson-Siegel (NS) model, introduced by \cite{NelsonSiegel87}, is the most popular empirical framework employed within the literature. \cite{diebold2006forecasting} generalised this model by introducing smooth dynamics for the latent factors, resulting in the dynamic NS (DNS) model, which outperforms the standard NS approach in an out-of-sample forecasting context. However, these approaches do not account for the potential impact of macroeconomic variables on the yield curve. To address this limitation, \cite{diebold2006macroeconomy} and \cite{coroneo2016unspanned} analysed the dynamic linkages between yield curve factors and macroeconomic factors in the US. \cite{Koopman2013} found evidence of interdependence between macroeconomic and latent factors in the US term structure, thus supporting the use of macroeconomic variables in combination with financial yield curve factors. Continuing along this line of research, \cite{bianchi2009great} applied a macro-factors augmented DNS model with time-varying parameters to UK data. Recently, \cite{Han2021} introduced a time-varying decay parameter for the arbitrage-free NS model and demonstrated that the relative factor loading reaches its peak just before starting the decline right before a recession.




It is worth emphasising that the literature on the term structure and the analysis of the relationship between yield curve factors and macroeconomic variables so far has been concerned with conditional mean models. However, the in-sample and forecasting performances of this approach are only sufficient in relation to the \textit{mean} of the yields, as the method is not suited to investigate the complete distribution of the yields.
To address the limitations of standard conditional mean Nelson-Siegel models and capture the complete picture of the conditional distribution of the yields, we propose a novel quantile regression Nelson-Siegel model with time-varying parameters (TVP-QR-NS) model.

\begin{figure}[H]
\includegraphics[width=1\textwidth]{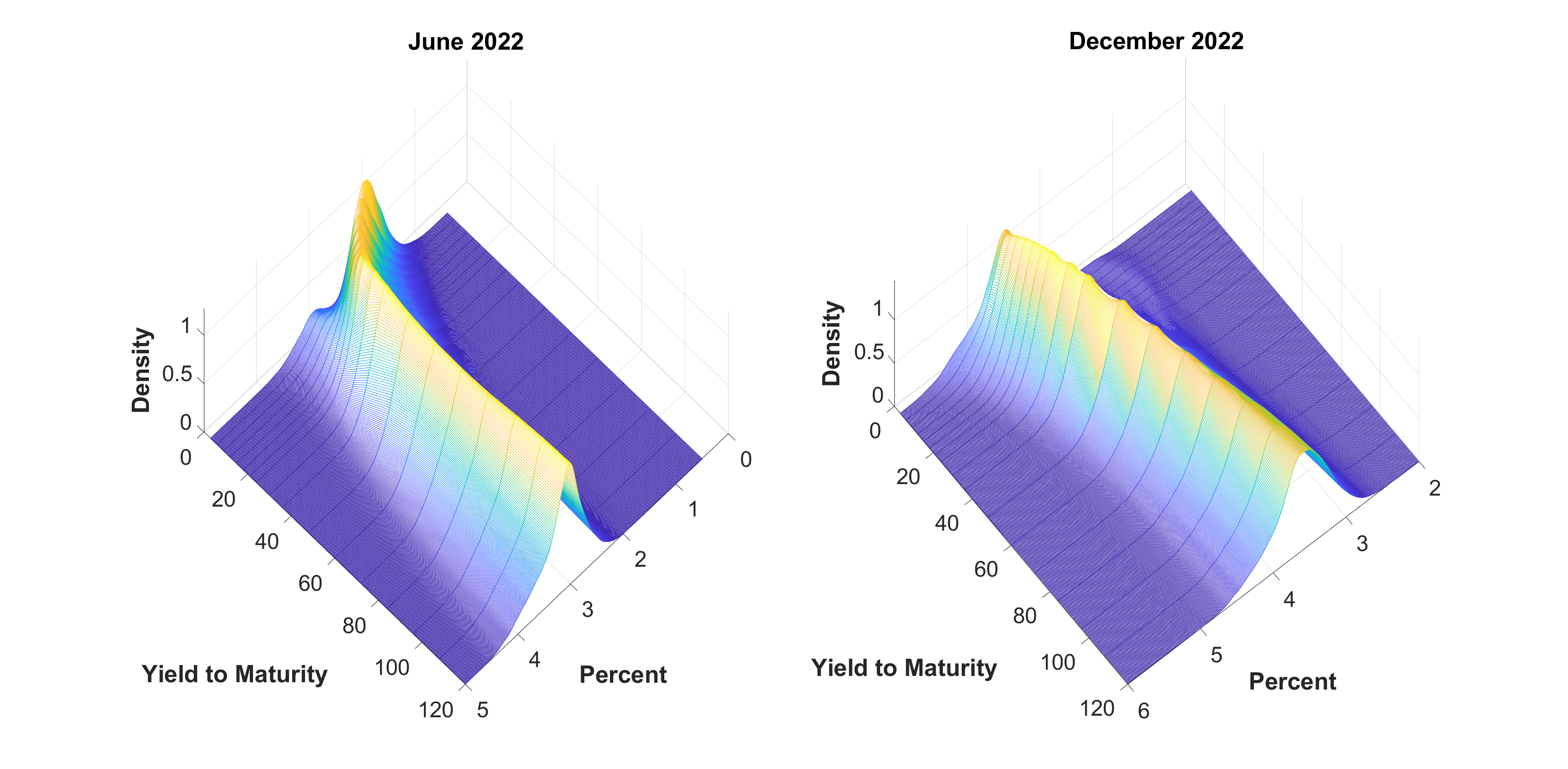}
\caption{Distributional term structure of the interest rate for June (left) and December (right) 2022.}
\label{fig:Density}
\end{figure}

We contribute to this literature by proposing a quantile version of the Nelson-Siegel model for the yield curve. Specifically, we extend the \cite{diebold2006macroeconomy} 'Yields-Macro' model to a quantile framework. Our approach assumes that the conditional $\tau$-quantile of the yields is given by the Nelson-Siegel three-factor specification, where the factors are (smoothly) time-varying. 
Quantile regression \citep[QR, see][]{koenker1978regression} does not assume a parametric likelihood for the conditional distribution of the response variable.
Unlike standard models based on the conditional mean and relying on symmetric second-moment dynamics, our novel framework offers robust modelling of the conditional quantiles, thus enabling a comprehensive investigation of the entire conditional distribution.
Moreover, by considering multiple quantile levels, we can follow the method proposed by \cite{mitchell2024constructing} and combine the quantiles to obtain the distributional yield curve at each point in time, as shown in Figure~\ref{fig:Density}. This permits the investigation of dynamic asymmetry within the distribution of the term structure of interest rates. This feature is particularly crucial when undertaking the modelling and forecasting of interest rate term structures, as emphasised in our empirical application.

Inference is performed within a Bayesian framework by leveraging the connection between a likelihood formed by the asymmetric Laplace distribution \citep{kotz2001laplace} and quantile regression provided by \cite{yu2001bayesian}, and \cite{petrella2019joint}.

The main feature of the proposed TVP-QR-NS model is the coupling of quantile regression with time-varying three-factor Nelson-Siegel specification.
In addition, to cope with the potentially high dimensionality of the latent dynamic factors, we use a precision sampler \citep{chan2009efficient,chan2020reducing} that allows us to sample the entire path of the factors jointly without any loop. Overall, this combination provides a simple framework and computationally efficient method to investigate and forecast the term structure at every quantile level of interest. Moreover, our novel econometric framework shares similarities with methodologies put forth by \cite{korobilis2024probabilistic, korobilis2024monitoring}. In the former, the authors propose a variational Bayes approach to estimate a quantile static factor model, while the latter extends the factor-augmented vector autoregression to a quantile framework. The primary distinction between our study and both of these lies in our specific focus on analysing the distribution of the yield curve. In contrast, their emphasis is on examining macroeconomic uncertainty within the US and macroeconomic risk across multiple countries.

The proposed method is applied to investigate and forecast the US term structure of interest rates, with particular emphasis on capturing the full distributional dynamics of key yield curve factors. To validate its utility, we conduct an out-of-sample forecasting exercise to evaluate the performance of the TVP-QR-NS model against standard conditional mean benchmarks widely used in the empirical literature. This study contributes to a growing body of research that focuses on forecasting macroeconomic variables from a distributional perspective (see \cite{carriero2022nowcasting, clark2024investigating, lenza2025density}). The results indicate that the TVP-QR-NS model consistently delivers superior tail risk forecasts relative to conventional benchmarks, particularly at longer forecast horizons and for long-term maturities. Moreover, employing the model confidence set methodology proposed by \citet{hansen2011model}, we find that the TVP-QR-NS model is systematically retained within the confidence sets for longer maturities and long-term forecast horizons, while standard conditional mean models are frequently excluded, underscoring the robustness and reliability of our approach.

The in-sample analysis highlights the advantages of the quantile regression framework over the conventional conditional mean Nelson-Siegel model. Unlike mean-based frameworks, our method enables estimation of the entire distribution of the three Nelson-Siegel factors, as well as yields at selected maturities over time. This comprehensive characterization allows for a more detailed assessment of the effects of recent economic crises, specifically the Great Recession and the COVID-19 pandemic, on the term structure.

The results uncover critical insights that are not accessible through conditional mean models. Specifically, the impact of the Great Recession and COVID-19 on the distribution of short-term interest rates diverges substantially, with clear distributional shifts evident across the two crises. In contrast, the distribution of long-term rates during the Great Recession exhibited no significant shift in location, although an increase in left-skewness was observed.

Further analysis shows that, following the Great Recession, a pronounced location shift occurred across the short, medium, and long ends of the yield curve, with the 10-year maturity exhibiting increased left-skewness. Conversely, during the COVID-19 recession, the location shift was confined to the long end of the yield curve and was accompanied by heightened dispersion. Moreover, we find robust evidence of quantile-dependent dynamics in the relationship between yield curve factors and macroeconomic conditions, underscoring the importance of capturing heterogeneity across the distribution.

The remainder of this article is organised as follows.
Section~\ref{sec:methods} introduces and describes the novel framework, then Section~\ref{sec:inference} presents the Bayesian approach to inference. Section~\ref{sec:results} details the empirical application. Finally, Section~\ref{sec:conclusion} concludes.

\section{Methodology}  \label{sec:methods}

\subsection{Nelson-Siegel Yield Curve model}
The standard Nelson-Siegel (NS) model of the yield curve assumes that
\begin{equation}
    y_t(m) = \beta_1 + \beta_2 \left( \frac{1-\exp^{-\lambda m}}{\lambda m} \right) + \beta_3 \left( \frac{1-\exp^{-\lambda m}}{\lambda m}-\exp^{-\lambda m} \right),
\end{equation}
where $m$ denotes the time to maturity; that is, given a set of parameters, \textit{the yield curve} is a deterministic function of maturity. \cite{diebold2006forecasting} pioneered the stochastic modelling of the yield curve and suggested the linear model
\begin{equation}
    y_t(m) = \beta_{1t} + \beta_{2t} \left( \frac{1-\exp^{-\lambda m}}{\lambda m} \right) + \beta_{3t} \left( \frac{1-\exp^{-\lambda m}}{\lambda m}-\exp^{-\lambda m} \right) + \epsilon_t(m)
\end{equation}
where $\epsilon_t(m) \sim \mathcal{N}(0,\sigma^2(m))$ is an independent normal innovation. Therefore, given a set of time-varying parameters, the \textit{expected value of the curve} is a deterministic function of maturity.

\subsection{A Time-Varying Parameters Quantile NS Model}
Assume that $y_t(m)\in \mathcal{Y} \subset \R$ for all maturities $m \in \R_+$, and let $F_{t}(\cdot;m)$ denote the cumulative distribution function of $y_t(m)$.
We propose a new approach that alleviates the parametric assumptions of $y_t(m)$'s by modelling the quantile function of $F$,
\begin{equation*}
    Q_t^\tau(m) = \inf\left\{x\in \mathcal{Y}: F_t(x;m)\geq \tau\right\}
\end{equation*}
for each quantile level $\tau \in (0,1)$ as follows
\begin{align}
   Q_t^\tau(m) = \beta_{1t}(\tau) + \beta_{2t}(\tau) \left( \frac{1-e^{-\lambda(\tau) m}}{\lambda(\tau) m} \right) + \beta_{3t}(\tau) \left( \frac{1-e^{-\lambda(\tau) m}}{\lambda(\tau) m} -e^{-\lambda(\tau) m} \right),
\label{eq:quantile_model}
\end{align}
where $m$ is the maturity, $\beta_{t1}^\tau$, $\beta_{t2}^\tau$, and $\beta_{t3}^\tau$ denote the (unobserved) level, slope and curvature factors, and $\lambda(\tau) \in\R_+$ is a decay parameter. 
Thus, by directly modelling the quantile function, we allow the yield curve factors to behave differently across the quantile levels.

Consider a series of observations at specific maturities $m \in \{1,\ldots,M\}$ and denote  with $\by_t = (y_t(m_1),\ldots,y_t(m_M))'$ the vector of yields at different maturity. 
Let $\bx(m;\lambda(\tau)) = \big( 1, \frac{1-e^{-\lambda(\tau) m}}{\lambda(\tau) m},\frac{1-e^{-\lambda(\tau) m}}{\lambda(\tau) m}-e^{-\lambda(\tau) m} \big)'$, and $\bX(\lambda(\tau)) = [ \bx(m_1;\lambda(\tau)), \bx(m_2;\lambda(\tau)),$ $\ldots, \bx(m_M;\lambda(\tau))]' \in \R^{M\times 3}$.
Then, the conditional quantile model can be written as
\begin{equation}
    Q^\tau_t = \bX(\lambda(\tau)) \bbeta_t(\tau)
\end{equation}
such that $Q^\tau_t = (Q^\tau_t(m_1),\ldots,Q^\tau_t(m_M))'$ and $\bbeta_t(\tau) = (\beta_{1t}(\tau), \beta_{2t}(\tau), \beta_{3t}(\tau))'$.
Moreover, at each quantile level $\tau$, the unobserved factors are modelled jointly with a vector of macroeconomic (observed) variables $\bz_t \in\R^r$. Specifically, the vector of time-varying (unobserved) yield curve and (observed) macroeconomic factors $\bphi_t = (\bbeta_t'(\tau), \bz_t')'$ is assumed to follow a stationary VAR(1) process
\begin{equation}
    \bphi_t = \bgamma_0 + \Gamma \bphi_{t-1} + \boldeta_t,  \qquad  \boldeta_t \sim \mathcal{N}(\mathbf{0}, \Omega).
\label{eq:factor_VAR_model}
\end{equation}
Differently from \cite{coroneo2016unspanned} that proposed an approach related to our in the context of a conditional mean model for the yields, we are modelling the conditional quantile of $y_t(m)$ and rely on an alternative specification of the interactions between the yield and the macroeconomic factors.

To design a regression model for the yield such that the conditional $\tau$-quantile corresponds to eq.~\eqref{eq:quantile_model}, we adopt a likelihood-based approach and rely on the properties of the asymmetric Laplace distribution \citep{kotz2001laplace}.
In particular, for a given quantile $\tau$, we consider a regression model where the innovation follows an asymmetric Laplace distribution $\epsilon_{t}(m) \sim \mathcal{AL}(\mu,\sigma,\tau)$, with location $\mu\in \R$, scale $\sigma >0$, skewness parameter $\tau \in (0,1)$, with density
\begin{equation*}
    p(x | \mu,\sigma,\tau) = \frac{\tau(1-\tau)}{\sigma} \exp\Big\{ -\rho_\tau\Big( \frac{x-\mu}{\sigma} \Big) \Big\},
\end{equation*}
where $\rho_\tau(x) = x(\tau-\I(x \leq \tau))$ is the check-loss function.
Therefore, combining the conditional quantile in eq.~\eqref{eq:quantile_model} with the asymmetric Laplace innovation results in
\begin{align}
    y_t(m) = \bx(m; \lambda(\tau))' \bbeta_t(\tau) + \epsilon_{t}(m), \qquad \epsilon_{t}(m) \sim \mathcal{AL}(0, \sigma, \tau).
\label{eq:QRNS_observation}
\end{align}
It is important to highlight that the asymmetric Laplace distribution admits a stochastic representation as a location-scale mixture of Gaussian \citep{kotz2001laplace}, thus allowing to rewrite eq.~\eqref{eq:QRNS_observation} as
\begin{align}
    y_t(m) \stackrel{d}{=} \bx(m;\lambda(\tau))' \bbeta_t(\tau) + w_{t}(m) \sigma \theta_{\tau,1}  + \sqrt{w_{t}(m)} \sigma \theta_{\tau,2} u_{t}(m), \quad u_{t}(m) \sim \mathcal{N}(0,1),
\label{eq:QRNS_observation_conditinoal}
\end{align}
where $w_{t}(m) \sim \mathcal{E}xp(1)$ is exponential with rate one. To establish an equivalence between the mode of the asymmetric Laplace distribution and the optimisation of the standard quantile regression problem in \cite{koenker1978regression}, the parameters $\theta_{\tau,1}, \theta_{\tau,2}$ are constrained to the following values \citep{kotz2001laplace}:
\begin{equation}
    \theta_{\tau,1} = \frac{1-2\tau}{\tau(1-\tau)}, \qquad
    \theta_{\tau,2} = \sqrt{\frac{2}{\tau(1-\tau)}}.
\end{equation}
In the existing literature, the decay parameter is often fixed at $\lambda = 0.0609$ \citep[e.g.,][]{diebold2006forecasting}. In contrast, we adopt an agnostic approach by estimating $\lambda(\tau)$ separately for each quantile, and assume a uniform prior distribution, $\lambda(\tau) \sim \mathcal{U}(0, 0.3)$ (see the appendix for further details). Given that the conditional posterior distribution of $\lambda(\tau)$ is non-standard, we employ a Griddy-Gibbs sampling step to draw from its posterior. Finally, the scale of the innovation is set to $\sigma=1.0$, as in the original quantile regression model \citep{yu2001bayesian}.

The combination of eq.~\eqref{eq:factor_VAR_model} and \eqref{eq:QRNS_observation} provides the state-space form of the proposed time-varying parameters quantile Nelson-Siegel model (TVP-QR-NS) as
\begin{equation}
\begin{split}
    y_t(m) & = \bx(m; \lambda(\tau))' \bbeta_t(\tau) + \epsilon_{t}(m),   \qquad   \epsilon_{t}(m) \sim \mathcal{AL}(0, \sigma, \tau) \\
    \bphi_t & = \bgamma_0 + \Gamma \bphi_{t-1} + \boldeta_t,  \qquad  \boldeta_t \sim \mathcal{N}(\mathbf{0}, \Omega),
\end{split}
\label{eq:TVP_QRNS_model}
\end{equation}

Denoting with $\bY = (\by_1,\ldots,\by_T)' \in\R^{T\times M}$ the matrix of observations, the joint density of the observables and auxiliary variables (conditional on the time-varying factors) is
\begin{align}
    p(\bY, \bw | \bbeta) & = \prod_{m=1}^M \mathcal{E}xp(w_{t}(m) | 1) \times \mathcal{N}\big( y_t(m) | \bx(m; \lambda(\tau))' \bbeta_t(\tau) + w_{t}(m) \sigma \theta_{\tau,1}, \: w_{t}(m) \sigma^2 \theta_{\tau,2}^2 \big).
\label{eq:complete_likelihood}
\end{align}

\section{Estimation}     \label{sec:inference}

This section is devoted to the prior description and the derivation of the precision sampler algorithm.

\subsection{Prior Specification}
We consider a multivariate Gaussian prior for the initial value of the time-varying factors:
\begin{equation}
    \bbeta_0(\tau) \sim \mathcal{N}(\underline{\bmu}_\beta, \underline{\Sigma}_\beta).
\end{equation}
By leveraging the mixture representation of the asymmetric Laplace distribution, we obtain a conditionally linear Gaussian state space for the latent dynamic vector $\bbeta_t$. Thus, we sample the full conditional posterior distribution of $\bbeta_t$ directly in a single loop, using a precision sampler \citep[e.g., see][]{chan2009efficient,chan2020reducing} as described in the following subsection.

Let us define $e_{t}(m) = y_t(m) - \bx(m;\lambda(\tau))' \bbeta_t$. Then, the full conditional posterior distribution of the auxiliary variable $w_{t}(m)$ is obtained as
\begin{align}
    w_{t}(m) | y_t(m), \bbeta_t 
    & \propto w_{t}(m)^{-\frac{1}{2}} \exp\left\{ -\frac{1}{2} \left[ \frac{\theta_{\tau,1}^2}{\theta_{\tau,2}^2} w_{t}(m) + 2 w_{t}(m)  +  w_{t}(m)^{-1} \frac{e_{t}(m)^2}{\theta_{\tau,2}^2 \sigma^2} \right] \right\},
\end{align}
which is the kernel of the generalized inverse Gaussian distribution $\text{GiG}(\overline{p}_w,\overline{a}_w,\overline{b}_{w,m,t})$ with
\begin{align*}
    \overline{p}_{w} = 1-\frac{1}{2}, \qquad
    \overline{a}_{w} & = 2 + \frac{\theta_{\tau,1}^2}{\theta_{\tau,2}^2}, \qquad
    \overline{b}_{w,m,t}  = \frac{e_{t}(m)^2}{\theta_{\tau,2}^2 \sigma^2}.
\end{align*}

Finally, we specify a Gaussian prior for the vectorised coefficient matrix of the state equation, $\bgamma = (\bgamma_0',\operatorname{vec}(\Gamma)')'$, and an inverse Wishart prior for the innovation covariance, that is
\begin{equation}
    \bgamma \sim \mathcal{N}(\underline{\bmu}_\Gamma, \underline{\Sigma}_\Gamma), \qquad
    \Omega \sim \mathcal{IW}(\underline{v}, \underline{V}).
\end{equation}
As the prior distributions are conjugate, the full conditional posteriors are straightforward to obtain as
\begin{equation}
    \bgamma | \bphi, \Omega \sim \mathcal{N}(\overline{\bmu}_\Gamma, \overline{\Sigma}_\Gamma), \qquad
    \Omega | \bphi, \gamma_0, \Gamma \sim \mathcal{IW}(\overline{v}, \overline{V}).
\end{equation}
The derivation of these full conditional posteriors is standard \citep[e.g, see][]{chan2020large}.

\subsection{The precision sampler}
While the conditional distribution of the parameters given the data and the set of unobserved latent factors is relatively standard, the most significant computational costs arise from sampling the large number of latent factors, denoted as $\bbeta_t(\tau)$.

Let us start by rewriting the measurement equation in \eqref{eq:TVP_QRNS_model} in compact form by stacking all observations together. Specifically, we define the stacked vectors $\by = (\by_1',\ldots,\by_T')'$, $\bbeta = (\bbeta_1(\tau)',\ldots,\bbeta_T(\tau)')'$, $\bw = (w_1,\ldots,w_T)'$, and $\bu = (u_1,\ldots,u_T)'$, we have
\begin{equation}
    \by = \left( \Id_T \otimes \bX(\lambda(\tau)) \right) \bbeta + \theta_{\tau,1} \sigma \bw + \bu,
\end{equation}
where $\bu \sim \mathcal{N}(\mathbf{0}_{TM}, \sigma^2 \theta_{\tau,2} \Omega_{1})$ with $\Omega_1 = \operatorname{diag}(\bw)$.
Similarly, we define $\bphi = (\bphi_1',\ldots,\bphi_T')'$ and rewrite the state equation in a matrix form as
\begin{equation}
    \bH \bphi = \tilde{\bgamma} + \boldeta, \qquad \boldeta \sim \mathcal{N}\left( \mathbf{0}_{T(r+3)}, \Omega \otimes \Id_T \right),
\end{equation}
where
\begin{equation*}
    \bH = \begin{pmatrix}
    \Id_{r+3} & \mathbf{0}_{r+3} & \mathbf{0}_{r+3} & \mathbf{0}_{r+3} & ... & \mathbf{0}_{r+3} & \mathbf{0}_{r+3} \\
    \mathbf{0}_{r+3} & -\Gamma&\Id_{r+3} & \mathbf{0}_{r+3} & ... & \mathbf{0}_{r+3} & \mathbf{0}_{r+3} \\
    ...&...&...&...&...&...&... \\
    ...&...&...&...&...& \Id_{r+3} & \mathbf{0}_{r+3} \\
    ...&...&...&...&...& -\Gamma & \Id_{r+3}
    \end{pmatrix}, \qquad
    \tilde{\bgamma} = \begin{pmatrix}
    \bgamma_0 \\
    \bgamma_0 \\
    ... \\
    \bgamma_0 \\
    \bgamma_0
    \end{pmatrix}.
\end{equation*}
Note that the unobserved factors are placed first in the VAR vector such that
\begin{equation*}
    \bphi = S_1 \bbeta + S_2 \bu,
\end{equation*}
where 
\begin{equation*}
    S_1 = \Id_T \otimes (\Id_3, \mathbf{0}_{3\times r})', \quad S_2 = \Id_T \otimes (\mathbf{0}_{3\times r}, \Id_{r})'.
\end{equation*}
Thus, we obtain the conditional posterior of the time-varying yield curve factors as $\bbeta | \bY, \bw, \Gamma, \bgamma_0, \Omega \propto \mathcal{N}\left( \bmu_b, K_b^{-1} \right)$, where 
\begin{align*}
    K_b & = \frac{1}{\sigma^2\theta_{\tau,2}} \big( \mathbf{I}_T\otimes \bX(\lambda(\tau)) \big)'\Omega_1^{-1} \left(\mathbf{I}_T \otimes \bX(\lambda) \right) + S_1\mathbf{H}' \big(\Id_T \otimes \Omega^{-1}\big) \mathbf{H}S_1, \\
    \bmu_b & = K_b^{-1} \Big( \big( \mathbf{I}_T \otimes \bX(\lambda(\tau)) \big)' \Omega_1^{-1} (\by-\theta_{\tau_1} \sigma \bw) + S_1 \mathbf{H}' \big(\Id_T \otimes \Omega^{-1}\big) \left( \bgamma-\mathbf{H}S_2 z \right) \Big).
\end{align*}

\subsection{Monotonicity}

Traditional estimation of quantile regression, including our strategy, involves fitting the quantile NS model separately at each quantile level, treating each quantile as an independent modeling task. Although this approach provides flexibility and ease of implementation, it overlooks the intrinsic monotonicity that should characterize conditional quantile functions by construction. As a result, the estimated quantile curves may violate the logical ordering across quantiles, potentially leading to internal inconsistencies and undermining the interpretability of the distributional forecasts.

To ensure monotonicity across quantile levels $\tau \in \left\{ \tau_1,...,\tau_K\right\}$, it is essential to move beyond separate estimations and instead estimate the quantile NS models jointly across $\tau$. This joint estimation framework requires imposing explicit monotonicity constraints across quantiles,
\begin{equation*}
\bx\big( m;\lambda(\tau_k) \big)' \bbeta_t(\tau_k)\leq  \bx\big( m;\lambda(\tau_{k+1}) \big)' \bbeta_t(\tau_{k+1})
\end{equation*}
for $m=1,\ldots, M, k=1,\ldots, K$ and $t=1,\ldots T$, thus preserving the fundamental ordering structure inherent to conditional quantile functions. 
However, this approach introduces a large number of cross-quantile constraints, and the resulting optimization problem becomes highly complex and computationally burdensome, to the point where joint estimation under monotonicity constraints can be practically infeasible in its full form. This poses a major challenge for implementing a theoretically coherent yet computationally tractable distributional forecasting framework. To construct a coherent predictive density from the individually estimated quantile NS models, we adopt the re-ordering technique proposed by \cite{chernozhukov2010quantile}, a standard approach in the literature for addressing the issue of quantile crossing. 

In the context of the NS model, the yield curve is endowed with additional structural interpretation that can inform both estimation and inference.
Specifically, the level factor in the NS model captures the long-run component of interest rates and asymptotically corresponds to the yield at extremely long maturities—essentially representing the ultimate long-term rate.
Conversely, the sum of the level and slope factors determines the instantaneous (short-term) interest rate, reflecting the yield at the shortest maturity. Based on these insights, we can potentially impose the following monotonicity constraints
\begin{equation}
    \bbeta_{t,1}(\tau_k)\leq \bbeta_{t,1}(\tau_{k+1})
\label{eq:mon1}
\end{equation}
and
\begin{equation}
    \bbeta_{t,1}(\tau_k)+ \bbeta_{t,2}(\tau_k)\leq \bbeta_{t,1}(\tau_{k+1})+ \bbeta_{t,2}(\tau_{k+1}).
\label{eq:mon2}
\end{equation}
Instead of directly imposing these monotonicity constraints within a joint estimation framework—such as through constrained MCMC, which would be computationally intensive—we adopt a more practical post-processing strategy. Specifically, we formulate a quadratic programming problem that adjusts the unconstrained estimates of $\bbeta_{t,1}(\tau_k)$ and $\bbeta_{t,2}(\tau_k)$ across quantile levels. The objective is to find new values that remain as close as possible to the originally estimated posterior mean, $\widehat{\bbeta}_{t,1}$ and $\widehat{\bbeta}_{t,2}$, while satisfying the monotonicity constraints. This post-estimation correction ensures internal consistency and preserves the structural interpretation of the NS model, while avoiding the computational burden of imposing constraints during the sampling phase.

\section{Empirical Application}  \label{sec:results}

This section focuses on evaluating the performance of the proposed TVP-QR-NS model applied to zero-coupon US Treasury yields. Specifically, the analysis is divided into two parts: an out-of-sample forecasting exercise and an in-sample assessment.

\subsection{Data}

Our empirical investigation employs zero-coupon US Treasury yields as constructed by \cite{liu2021reconstructing} through a non-parametric kernel smoothing methodology. The application of this novel yield curve dataset by \cite{liu2021reconstructing} is demonstrated to offer a more precise depiction of the underlying data when compared to the alternative metric presented by \cite{gurkaynak2007us}. Following the methodology established by \cite{diebold2006macroeconomy}, our analysis encompasses zero-coupon US Treasury maturities spanning 3, 6, 9, 12, 15, 18, 21, 24, 30, 36, 48, 60, 72, 84, 96, 108, and 120 months. Additionally, our investigation includes key monthly macroeconomic indicators — Industrial Production, CPI Inflation, and the Fed Funds Rate — in line with the framework suggested by \cite{diebold2006macroeconomy}. These macroeconomic variables are chosen to capture the dynamic aspects of the macroeconomy and are all sourced from the US FRED database.  Both Industrial Production and CPI Inflation were transformed to log-differenced growth rates. Our study period extends from November 1971 to December 2024.

\FloatBarrier
\subsection{Forecasting Yields Tail Risk}

To validate the utility of our proposed framework, we conduct a pseudo-out-of-sample forecasting exercise to evaluate the performance of the quantile Nelson-Siegel approach in capturing tail risks in zero-coupon yields. The benchmark model employed for comparison is a univariate autoregressive model with one lag. Our proposed model’s performance is assessed relative to this baseline as well as three alternative conditional mean models, as detailed in Table~\ref{tab:Models}. These competing specifications include: (i) a time-varying parameter model with stochastic volatility (SV); (ii) the large Bayesian vector autoregression (VAR) with common stochastic volatility (CSV) developed by \cite{chan2020large}; and (iii) the conditional mean formulation of the dynamic Nelson-Siegel model of \cite{diebold2006macroeconomy}.

\begin{table}[H]
\centering
\caption{Overview of Competing Models}
\label{tab:Models}
\begin{tabular}{ll}
\toprule
\textbf{Model} & \textbf{Description} \\
\midrule
AR(1) & Autoregressive model with one lag (Benchmark model). \\
TVP-AR(1)-SV & Time-varying parameter autoregressive model with SV and one lag. \\
BVAR(1)-CSV & Large Bayesian VAR with one lag and common stochastic volatility. \\
DNS & Dynamic Nelson-Siegel model of \cite{diebold2006macroeconomy}. \\
TVP-QNS & Proposed Quantile-based time-varying Nelson-Siegel model. \\
\bottomrule
\end{tabular}
\end{table}

The initial holdout period spans from November 1971 to January 2000, with a subsequent forecast evaluation period extending from February 2000 to December 2024. We generate 1st, 12th, and 36th month-ahead tail risk forecasts and evaluate the predictive performance of the competing models using the quantile score (QS) \citep[see][]{giacomini2005evaluation, carriero2022nowcasting}. Following \cite{gneiting2011comparing}, we define the QS as:
\begin{align}
    QS_{\tau,i,t} = (y_{t}(m) - \mathcal{Q}_{\tau,i,t}) - (\tau - \mathbb{I} \{y_{t}(m) \leq \mathcal{Q}_{\tau,i,t}\}).
\label{eq:QS}
\end{align}
Here, $\mathcal{Q}_{\tau,i,t}$ denotes the predictive quantile of the $m$-th yield, and $\mathbb{I} \{y_{t}(m) \leq \mathcal{Q}_{\tau,i,t}\}$ takes the value 1 if the realised value is at or below the predictive quantile and 0 otherwise. We evaluate the QS in the upper and lower tails by setting $\tau=0.9$ and $\tau=0.1$, respectively.

\begin{table}[H]
\caption{Average 10\% (Lower-Tail) Quantile Scores for Selected Maturities Relative to an AR(1) Benchmark}
\centering
\label{tab:lower_tail_scores}
\resizebox{\textwidth}{!}{%
\begin{tabular}{lccccccccccccccc}
\toprule
\textbf{Model} & \textbf{3m} & $p_{\text{MCS}}$ & \textbf{6m} & $p_{\text{MCS}}$ & \textbf{12m} & $p_{\text{MCS}}$ & \textbf{24m} & $p_{\text{MCS}}$ & \textbf{36m} & $p_{\text{MCS}}$ & \textbf{60m} & $p_{\text{MCS}}$ & \textbf{120m} & $p_{\text{MCS}}$ & \textbf{Avg.} \\
\midrule
\multicolumn{16}{c}{\textit{1 Month-Ahead Forecast}} \\
\midrule
TVP-AR(1)-SV   & 0.43 & 0.00 & 0.42 & 0.01 & 0.50 & 0.03 & 0.63 & 0.02 & 0.72 & 0.00 & 0.81 & 0.00 & 0.90 & 0.00 & 0.68 \\
BVAR(1)-CSV   & 0.28 & 1.00{*}{*} & 0.31 & 1.00{*}{*} & 0.38 & 1.00{*}{*} & 0.50 & 1.00{*}{*} & 0.57 & 0.64{*}{*} & 0.63 & 0.23{*} & 0.65 & 1.00{*}{*} & 0.52 \\
DNS           & 1.72 & 0.00 & 1.32 & 0.00 & 1.06 & 0.00 & 1.13 & 0.00 & 1.23 & 0.00 & 1.29 & 0.00 & 1.07 & 0.00 & 1.20 \\
TVP-QNS       & 0.48 & 0.00 & 0.46 & 0.00 & 0.47 & 0.04 & 0.53 & 0.46{*}{*} & 0.55 & 1.00{*}{*} & 0.57 & 1.00{*}{*} & 0.70 & 0.41{*}{*} & \textbf{0.55} \\
\midrule
\multicolumn{16}{c}{\textit{12 Month-Ahead Forecast}} \\
\midrule
TVP-AR(1)-SV  & 1.60 & 0.00 & 1.27 & 0.00 & 1.39 & 0.00 & 1.27 & 0.00 & 1.07 & 0.00 & 0.94 & 0.00 & 0.88 & 0.00 & 1.15 \\
BVAR(1)-CSV   & 0.73 & 0.18{*} & 0.77 & 0.18{*} & 0.82 & 0.17{*} & 0.90 & 0.18{*} & 0.93 & 0.18{*} & 0.94 & 0.18{*} & 0.91 & 0.18{*} & 0.88 \\
DNS            & 1.15 & 0.00 & 1.10 & 0.00 & 1.05 & 0.00 & 1.08 & 0.00 & 1.11 & 0.00 & 1.13 & 0.00 & 1.04 & 0.00 & 1.09 \\
TVP-QNS        & 0.76 & 1.00{*}{*} & 0.75 & 1.00{*}{*} & 0.73 & 1.00{*}{*} & 0.76 & 1.00{*}{*} & 0.76 & 1.00{*}{*} & 0.74 & 1.00{*}{*} & 0.64 & 1.00{*}{*} & \textbf{0.73} \\
\midrule
\multicolumn{16}{c}{\textit{36 Month-Ahead Forecast}} \\
\midrule
TVP-AR(1)-SV   & -- & 0.00 & -- & 0.00 & -- & 0.00 & -- & 0.00 & -- & 0.00 & -- & 0.00 & 2.84 & 0.00 & -- \\
BVAR(1)-CSV    & 1.13 & 0.00 & 1.17 & 0.00 & 1.19 & 0.00 & 1.24 & 0.00 & 1.26 & 0.00 & 1.27 & 0.00 & 1.25 & 0.00 & 1.23 \\
DNS            & 1.04 & 0.00 & 1.03 & 0.00 & 1.01 & 0.00 & 1.05 & 0.00 & 1.08 & 0.00 & 1.12 & 0.00 & 1.10 & 0.00 & 1.07 \\
TVP-QNS       & 0.79 & 1.00{*}{*} & 0.79 & 1.00{*}{*} & 0.79 & 1.00{*}{*} & 0.84 & 1.00{*}{*} & 0.86 & 1.00{*}{*} & 0.87 & 1.00{*}{*} & 0.79 & 1.00{*}{*} & \textbf{0.83} \\
\bottomrule
\end{tabular}}
\begin{flushleft}
\footnotesize \textit{Notes:} Superscripts $^{*}$ and $^{**}$ indicate inclusion in the model confidence sets $\widehat{\mathcal{M}_{90\%}^{*}}$ and $\widehat{\mathcal{M}_{75\%}^{*}}$ respectively, as per \citet{hansen2011model}. The reported average score is computed across all 17 maturities. A dash ('–') indicates that the quantile scores for the TVP-AR(1)-SV model are exceptionally large and thus omitted from the table for clarity.
\end{flushleft}
\end{table}

Tables~\ref{tab:lower_tail_scores} and~\ref{tab:upper_tail_scores} present the 10th and 90th percentile quantile scores (QS) from the out-of-sample forecasting exercise for selected maturities. We also report the corresponding p-values from the Model Confidence Set (MCS) procedure proposed by \citet{hansen2011model}. The results suggest that the proposed quantile Nelson-Siegel model delivers superior tail risk forecasts, particularly at longer forecast horizons and for long-term maturities. At shorter horizons, however, the BVAR(1)-CSV model remains competitive and, in some cases, outperforms the proposed model in forecasting tail risks. Importantly, the quantile Nelson-Siegel model consistently outperforms its conditional mean counterparts, based on \citet{diebold2006macroeconomy}, across all forecast horizons and quantile levels. Moreover, the MCS analysis shows that the proposed model is always included in the $\widehat{\mathcal{M}_{90\%}^{*}}$ and $\widehat{\mathcal{M}_{75\%}^{*}}$ confidence sets, especially for longer maturities and extended forecast horizons. By contrast, the three conditional mean models are frequently excluded from these sets. Overall, these findings provide strong empirical support for the proposed time-varying parameter quantile Nelson-Siegel model as a robust tool for forecasting tail risk.

\begin{table}[H]
\caption{Average 90\% (Upper-Tail) Quantile Scores for Selected Maturities Relative to an AR(1) Benchmark}
\centering
\label{tab:upper_tail_scores}
\resizebox{\textwidth}{!}{%
\begin{tabular}{lccccccccccccccc}
\toprule
\textbf{Model} & \textbf{3m} & $p_{\text{MCS}}$ & \textbf{6m} & $p_{\text{MCS}}$ & \textbf{12m} & $p_{\text{MCS}}$ & \textbf{24m} & $p_{\text{MCS}}$ & \textbf{36m} & $p_{\text{MCS}}$ & \textbf{60m} & $p_{\text{MCS}}$ & \textbf{120m} & $p_{\text{MCS}}$ & \textbf{Avg.} \\
\midrule
\multicolumn{16}{c}{\textit{1 Month-Ahead Forecast}} \\
\midrule
TVP-AR(1)-SV   & 0.37 & 0.00 & 0.39 & 0.01 & 0.49 & 0.03 & 0.65 & 0.00 & 0.75 & 0.00 & 0.85 & 0.00 & 0.90 & 0.00 & 0.69\\
BVAR(1)-CSV    & 0.21 & 1.00{*}{*} & 0.24 & 1.00{*}{*} & 0.31 & 1.00{*}{*} & 0.40 & 1.00{*}{*} & 0.46 & 1.00{*}{*} & 0.54 & 1.00{*}{*} & 0.62 & 1.00{*}{*} & 0.44 \\
DNS           & 1.09 & 0.00 & 1.04 & 0.00 & 1.15 & 0.00 & 1.16 & 0.00 & 1.12 & 0.00 & 1.09 & 0.00 & 1.55 & 0.00 & 1.17 \\
TVP-QNS        & 0.53 & 0.00 & 0.57 & 0.00 & 0.63 & 0.04 & 0.59 & 0.00 & 0.56 & 0.02 & 0.56 & 0.70{*}{*} & 0.83 & 0.00 & \textbf{0.61} \\
\midrule
\multicolumn{16}{c}{\textit{12 Month-Ahead Forecast}} \\
\midrule
TVP-AR(1)-SV   & 0.89 & 1.00{*}{*} & 0.85 & 1.00{*}{*} & 0.90 & 1.00{*}{*} & 0.92 & 1.00{*}{*} & 0.91 & 1.00{*}{*} & 0.90 & 1.00{*}{*} & 0.87 & 1.00{*}{*} & 0.89 \\
BVAR(1)-CSV    & 0.83 & 0.93{*}{*} & 0.87 & 0.93{*}{*} & 0.93 & 0.93{*}{*} & 0.97 & 0.93{*}{*} & 0.99 & 0.94{*}{*} & 1.03 & 0.93{*}{*} & 1.09 & 0.94{*}{*} & 0.98 \\
DNS           & 1.07 & 0.00 & 1.08 & 0.00 & 1.10 & 0.00 & 1.07 & 0.00 & 1.03 & 0.00 & 0.98 & 0.00 & 1.06 & 0.00 & 1.04 \\
TVP-QNS        & 0.99 & 0.93{*}{*} & 0.99 & 0.93{*}{*} & 0.98 & 0.93{*}{*} & 0.90 & 0.93{*}{*} & 0.83 & 0.94{*}{*} & 0.76 & 0.93{*}{*} & 0.77 & 0.94{*}{*} & \textbf{0.86} \\
\midrule
\multicolumn{16}{c}{\textit{36 Month-Ahead Forecast}} \\
\midrule
TVP-AR(1)-SV   & -- & 0.00 & -- & 0.00 & -- & 0.00 & -- & 0.00 & -- & 0.00 & -- & 0.00 & 1.37 & 0.00 & -- \\
BVAR(1)-CSV     & 1.68 & 0.00 & 1.68 & 0.00 & 1.70 & 0.00 & 1.72 & 0.00 & 1.71 & 0.00 & 1.76 & 0.00 & 1.90 & 0.00 & 1.75 \\
DNS            & 1.07 & 0.05 & 1.06 & 0.26$^{**}$ & 1.06 & 0.74$^{**}$ & 1.03 & 0.05 & 0.99 & 0.00 & 0.96 & 0.00 & 1.03 & 0.00 & 1.02 \\
TVP-QNS        & 1.06 & 1.00{*}{*} & 1.03 & 1.00{*}{*} & 1.01 & 1.00{*}{*} & 0.95 & 1.00{*}{*} & 0.90 & 1.00{*}{*} & 0.86 & 1.00{*}{*} & 0.87 & 1.00{*}{*} & \textbf{0.93} \\
\bottomrule
\end{tabular}}
\begin{flushleft}
\footnotesize \textit{Notes:} Superscripts $^{*}$ and $^{**}$ indicate inclusion in the model confidence sets $\widehat{\mathcal{M}_{90\%}^{*}}$ and $\widehat{\mathcal{M}_{75\%}^{*}}$ respectively, as per \citet{hansen2011model}. The reported average score is computed across all 17 maturities. A dash ('–') indicates that the quantile scores for the TVP-AR(1)-SV model are exceptionally large and thus omitted from the table for clarity.
\end{flushleft}
\end{table}

\subsection{In-Sample Results}

Previous studies, including \cite{diebold2006forecasting, diebold2006macroeconomy, diebold2008global, mumtaz2009time, monch2012term}, focused on the yield curve and its effects within a conditional mean framework. Our study is distinct from the former as it is the first to explore the yield curve from a quantile or distributional perspective.

This section presents our empirical in-sample results. The first subsection reports posterior estimates of the yield curve factors and the decay parameter across various quantile levels. The second subsection investigates the distributional properties of short- and long-term interest rates, as well as selected Treasury maturities, during recent crises—the Great Recession and the COVID-19 pandemic. Finally, the third subsection analyses the relationship between yield curve factors and macroeconomic variables from a quantile perspective.

\subsubsection{Posterior Estimates of the Decay Parameter}

\begin{figure}[H]
\includegraphics[width=1\textwidth]{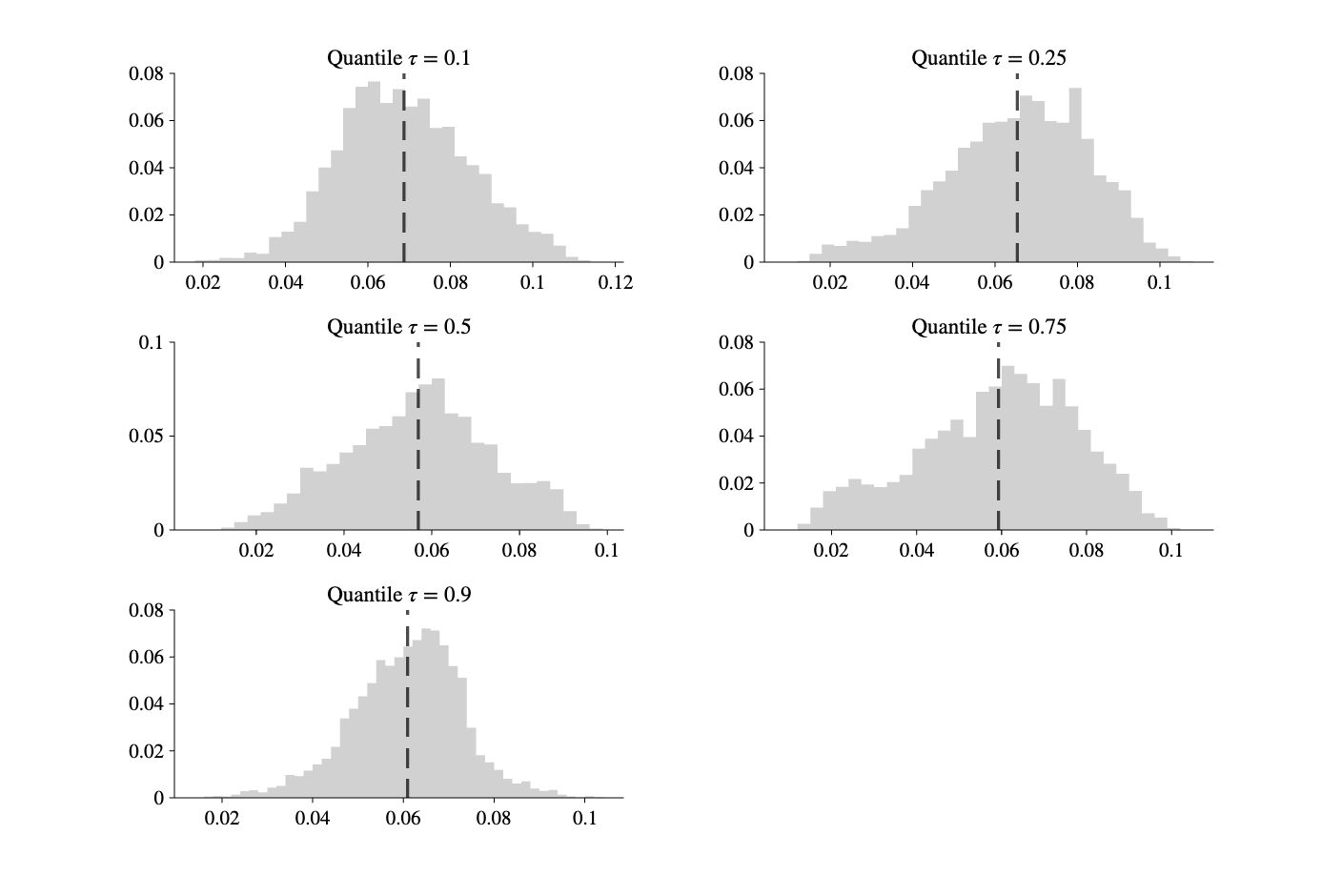}
\caption{Posterior distribution of $\lambda(\tau)$ for the 10\%, 25\%, 50\%, 75\% and 90\% quantiles, with the posterior mean (dashed line).}
\label{fig:lambda}
\end{figure}

Figure \ref{fig:lambda} presents the posterior distribution of the decay parameter $\lambda(\tau)$ across five selected quantiles (10\%, 25\%, 50\%, 75\%, and 90\%). The results indicate that, for each quantile, the posterior mass of $\lambda(\tau)$ is concentrated in the range of approximately 0.06 to 0.07. This is broadly consistent with the fixed value of $\lambda = 0.0609$ proposed by \citet{diebold2006macroeconomy}.

\subsubsection{Quantile Yield Curve Factor Estimates}

\begin{figure}[H]
\includegraphics[width=1\textwidth]{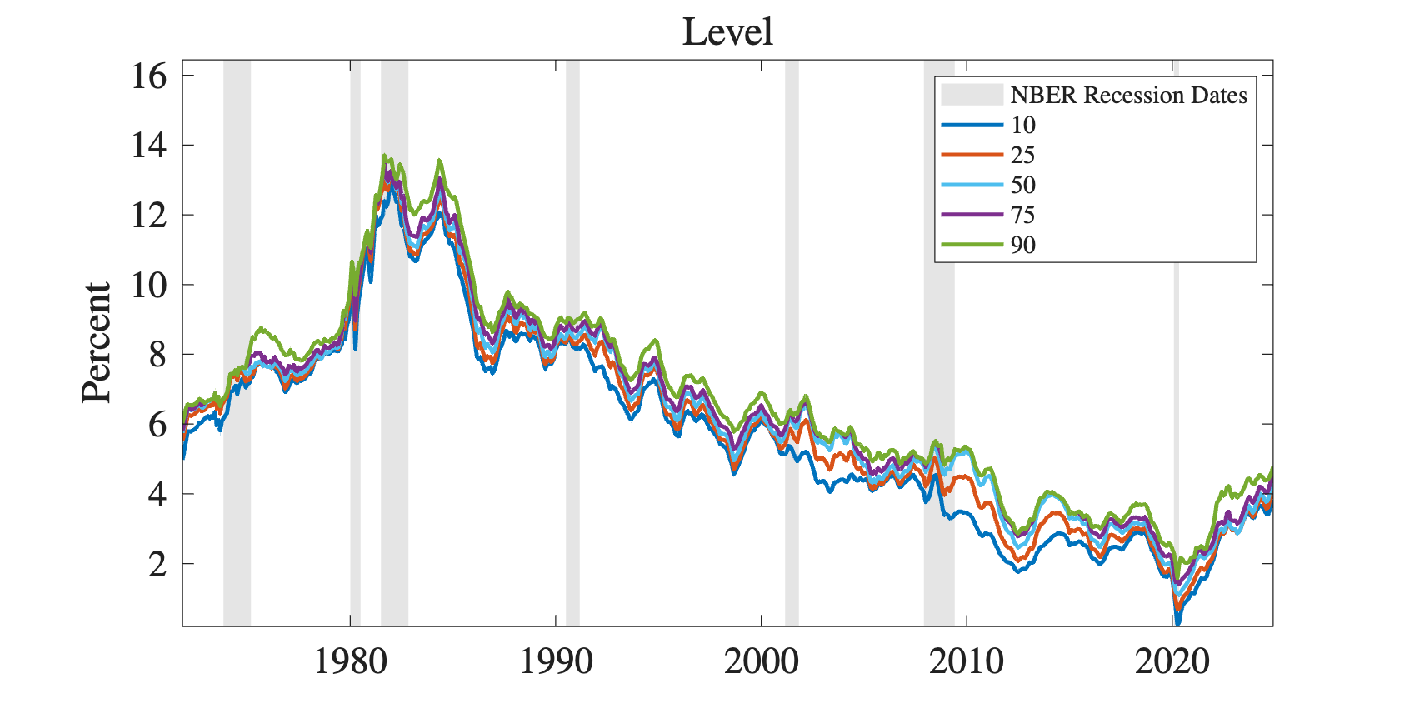}
\caption{Posterior mean estimates of the level factor for each of the five quantiles (10\%, 25\%, 50\%, 75\%, 90\%). NBER recession periods in grey shades.}
\label{fig:Levelmacro}
\end{figure}

\begin{figure}[H]
\includegraphics[width=1\textwidth]{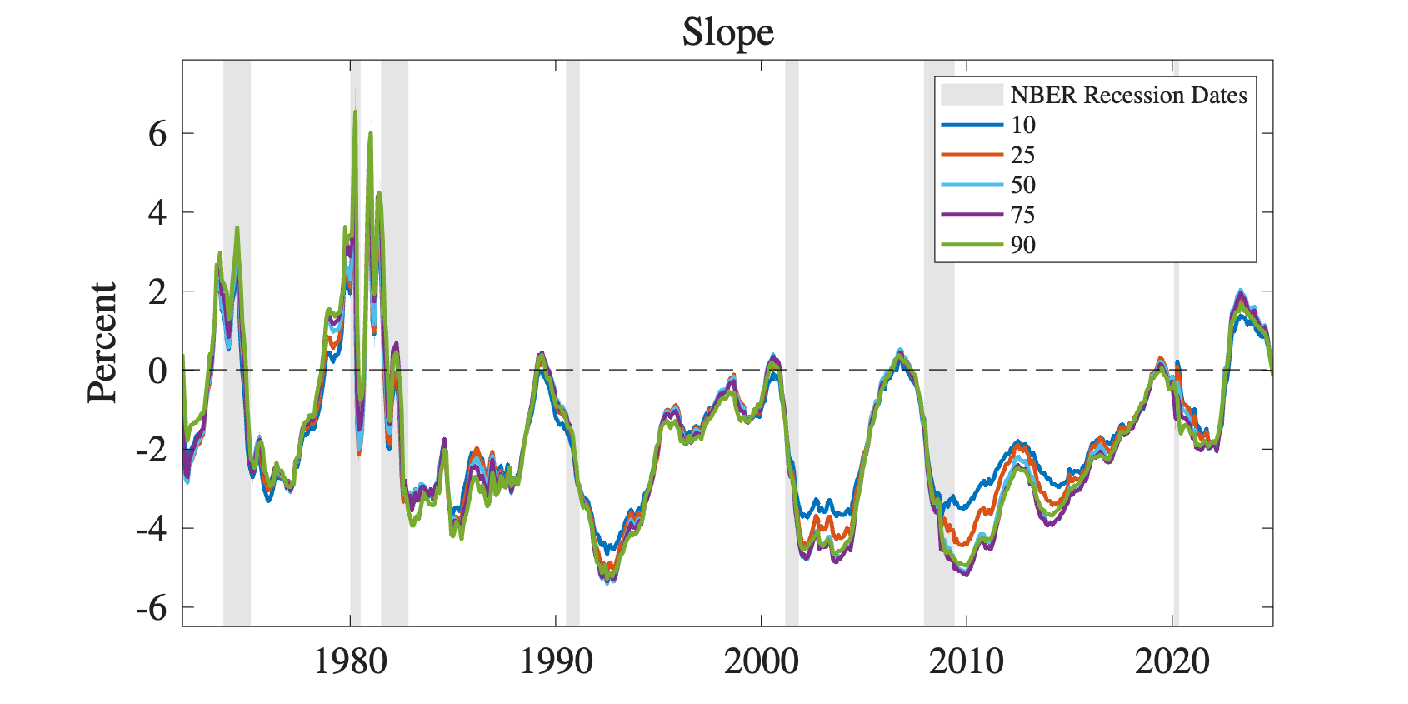}
\caption{Posterior mean estimates of the slope factor for each of the five quantiles (10\%, 25\%, 50\%, 75\%, 90\%). NBER recession periods in grey shades.}
\label{fig:Slopemacro}
\end{figure}

\begin{figure}[H]
\includegraphics[width=1\textwidth]{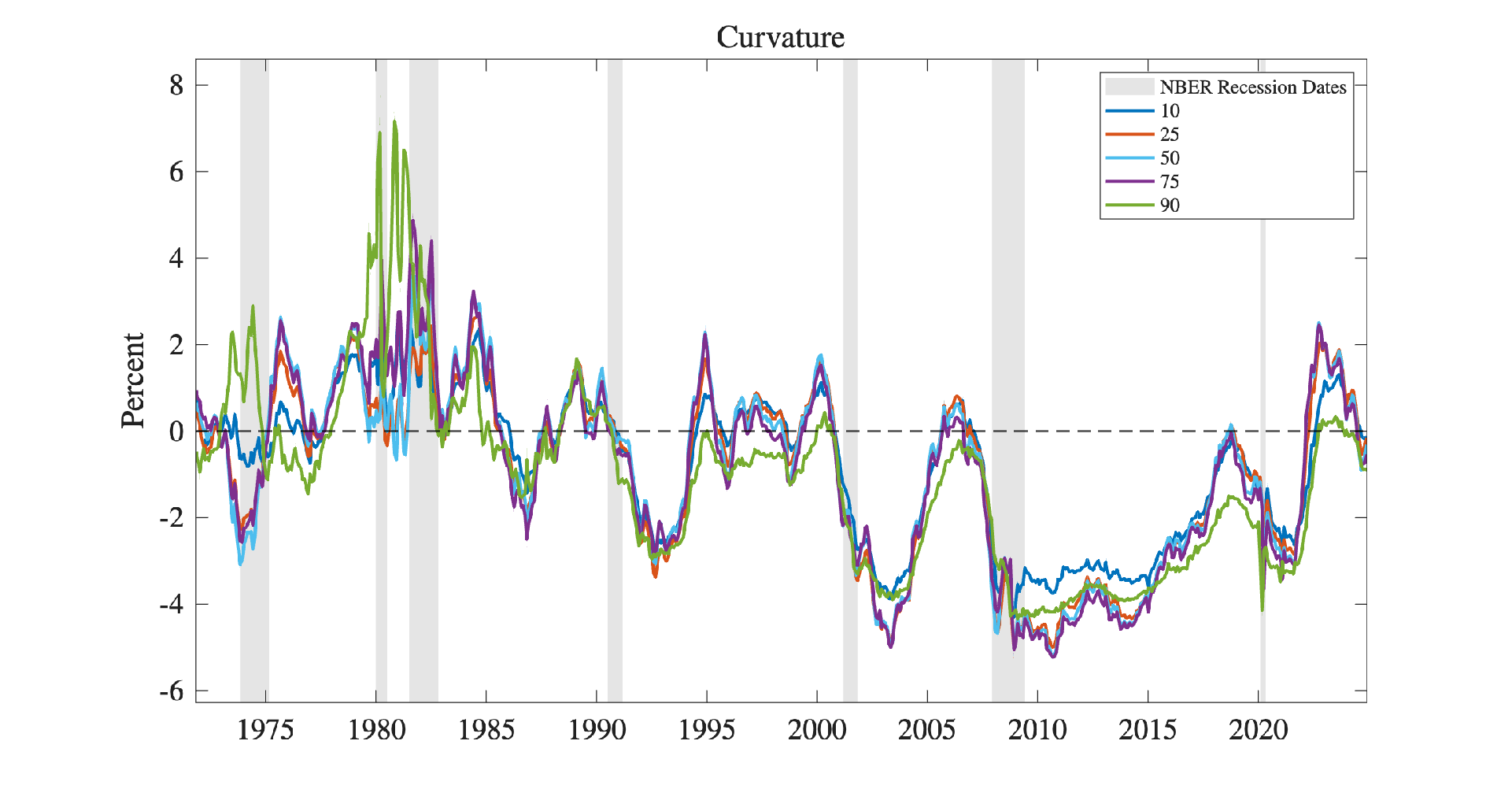}
\caption{Posterior mean estimates of the curvature factor for each of the five quantiles (10\%, 25\%, 50\%, 75\%, 90\%). NBER recession periods in grey shades.}
\label{fig:Curvmacro}
\end{figure}

Figures \ref{fig:Levelmacro}, \ref{fig:Slopemacro}, and \ref{fig:Curvmacro} depict the posterior mean estimates of time-varying yield curve factors at five quantile levels, providing insights into their dynamic evolution. A consistent finding across all graphs is the presence of substantial temporal variation. Beginning with the level factor, interpreted as the long-term interest rate (i.e., yield as maturity approaches infinity), a steady decline in this rate is discernible across all quantiles since the oil price crisis of the 1980s. Figure \ref{fig:Levelmacro} reveals a distinct pattern: following both the dot-com recession in the early 2000s and the Great Recession of 2008-09, the posterior estimates for the 90th and median quantiles closely align, while those for the 10th quantile diverge. Conversely, after the COVID-19 recession, the opposite trend emerged, with the 10th and median quantiles exhibiting proximity while the 90th quantile diverged.
This evidence suggests that post the early 2000s recession and the Great Recession, the distribution of the long-term interest rate skewed leftwards. In contrast, after the COVID-19 recession, the distribution skewed rightwards, likely attributed to the pronounced inflationary pressures experienced in the aftermath of the COVID-19 period.

Turning to the slope and curvature of the yield curve in Figures \ref{fig:Slopemacro} and \ref{fig:Curvmacro}, the degree of uncertainty associated with these factors varies over time. Notably, the curvature factor exhibits higher volatility across all quantiles than its slope counterpart.\footnote{We impose the monotonicity constraint specified in equation \eqref{eq:mon2} on the short-term interest rate, defined as the sum of the level and slope factors. Figure \ref{fig:short} in the appendix displays the posterior estimates of the short-term rate across the five quantiles, which exhibit no crossing and are thus consistent with the imposed constraint. However, this restriction does not imply monotonicity in the slope factor alone, which may still exhibit crossings across quantiles. In contrast, crossings in the conditional quantiles of the curvature factor are not of concern, as the curvature does not correspond directly to any specific yield maturity.} A distinct pattern is evident before 1985, where the posterior estimates for the slope factor across all five quantiles are closely clustered, indicating a period of low uncertainty. In contrast, during the oil price crisis of the 1980s, the curvature factor displayed substantial dispersion across quantiles.

Post-recessionary periods following the early 1990s, the 2000s, and the Great Recession of 2008-09 reveal elevated uncertainty in both slope and curvature factors. Specifically, convergence is observed between the median and the 90th quantile, with the 10th quantile exhibiting a discernible deviation from this convergent trajectory. In the aftermath of the COVID-19 recession, a reverse trajectory emerges, with the 90th quantile diverging from the 10th and median quantiles. The associated uncertainty surrounding both factors appears comparatively subdued compared to the preceding recessions of the 2000s.
Both the early 2000s recession and the Great Recession were attributed to financial crises. Our empirical findings suggest a potential connection between periods following financial crises and increased uncertainty in the slope and curvature of the yield curve.

It is worthwhile mentioning that these results could not be obtained using a conditional mean approach, which is instead able to produce estimates of the mean yield factors over time. Conversely, the proposed quantile approach enables deeper and more detailed analyses of the term structure while maintaining a low computational cost.

\subsubsection{Distributional Characteristics of Treasury Yields Across the Term Structure}

The proposed novel framework offers a distinct advantage by enabling the examination of the distributional properties of both short- and long-term interest rates, as well as selected yield maturities, over time. As previously discussed, the long-term interest rate is identified with the level factor, whereas the short-term interest rate reflects a combination of the level and slope factors. We follow the methodology outlined by \cite{mitchell2024constructing} to construct the empirical distributions of these rates and maturities. Specifically, we adopt an equal-weighting scheme to aggregate all MCMC draws across five quantiles and apply the Epanechnikov kernel, as proposed by \cite{gaglianone2012constructing}.

\begin{figure}[H]
\centering
\begin{tabular}{c}
\includegraphics[trim=0mm 0mm 0mm 0mm,clip,scale=0.44]{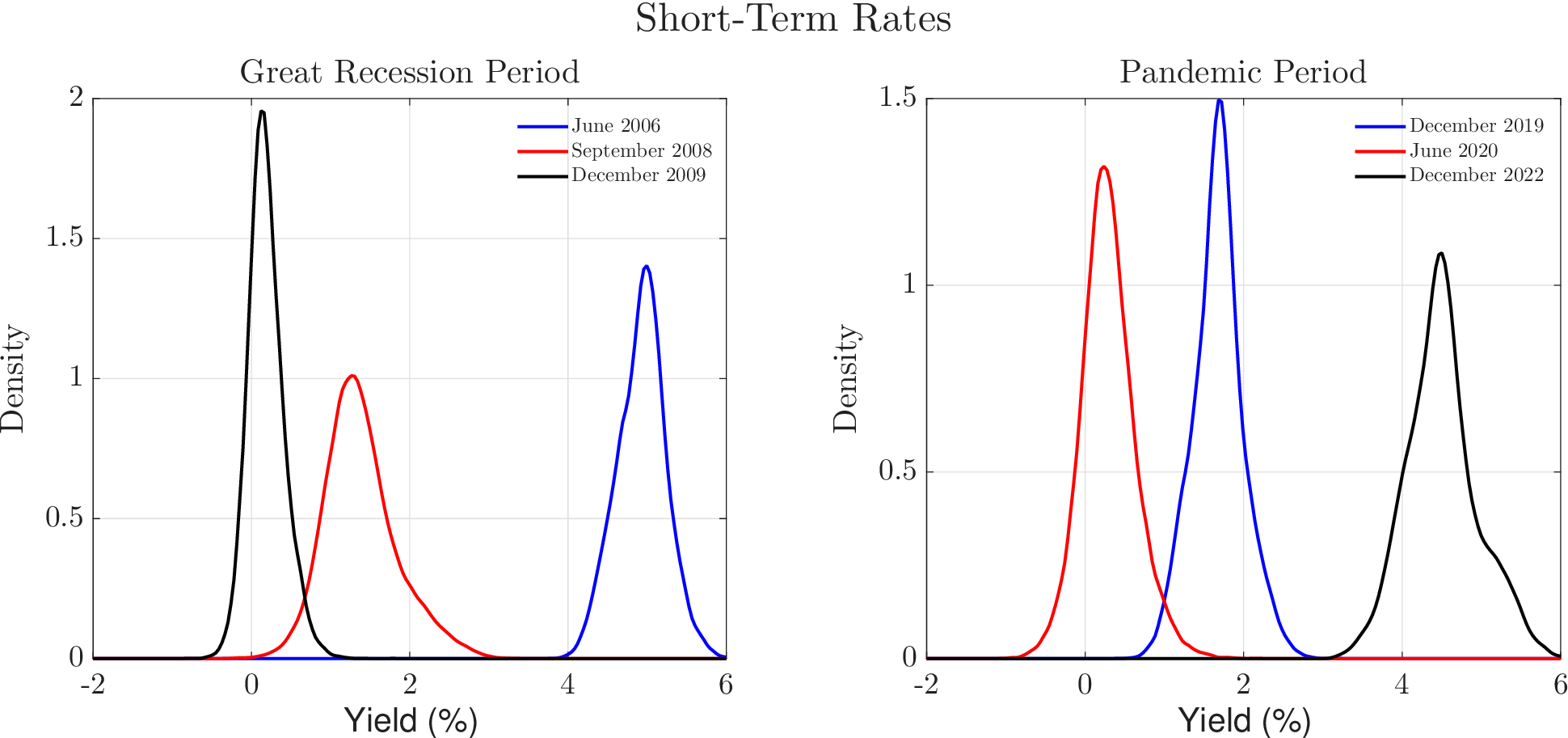} \\
\includegraphics[trim=0mm 0mm 0mm 0mm,clip,scale=0.44]{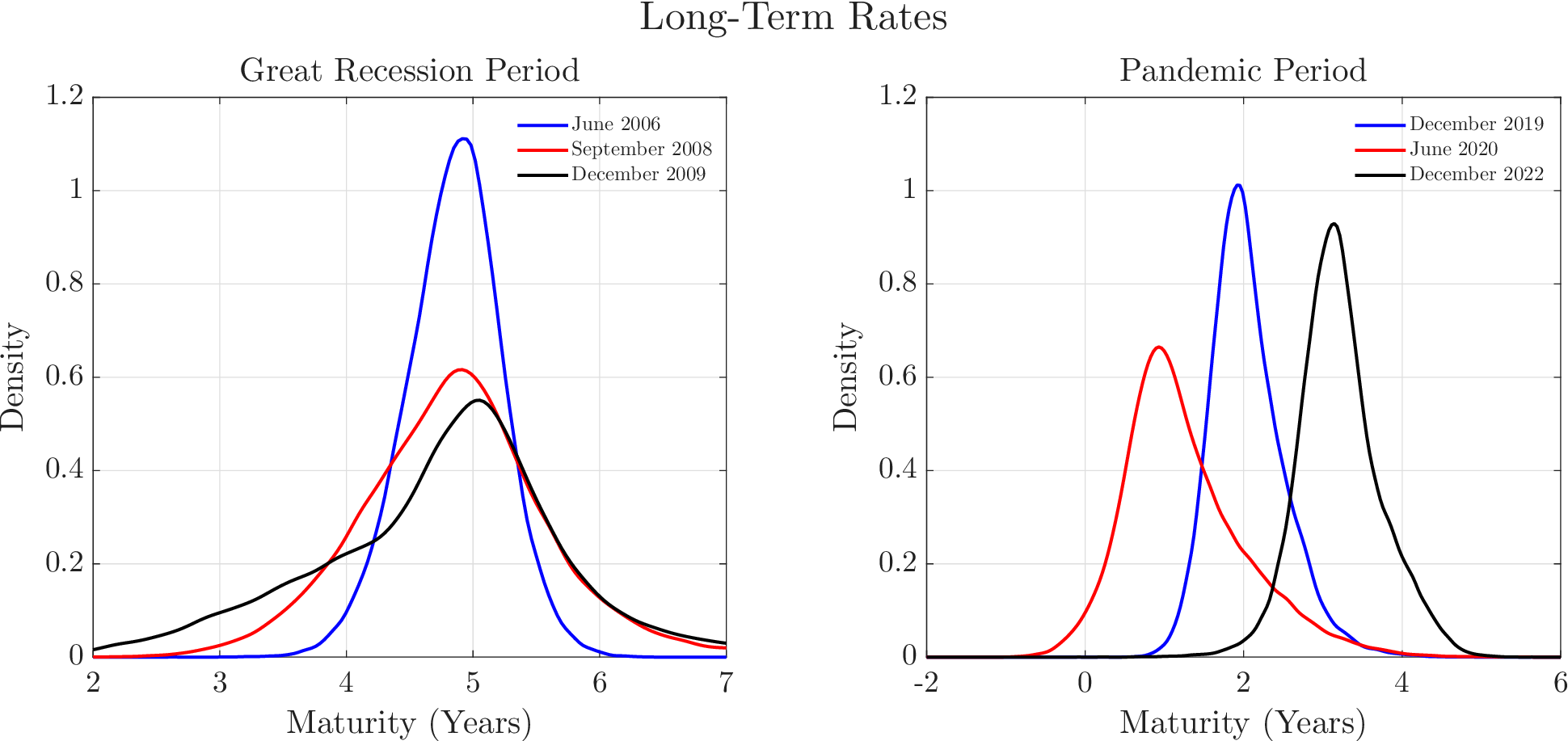}
\end{tabular}
\caption{Distribution of the short-term and long-term (level) factors during the Great Recession (left) and the COVID-19 pandemic (right).}
\label{fig:short_long_term_density}
\end{figure}

The previous analysis revealed an increase in the degree of dispersion among quantiles during post-recessionary periods, particularly in the 2000s. Expanding on this finding, we investigate whether the distributional features of short- and long-term interest rates exhibit temporal variations across the two most recent recessions. Figure \ref{fig:short_long_term_density} illustrates the distributions of both short- and long-term interest rates for selected months preceding and following the Great Recession and the COVID-19 recession.

Focusing first on the short-term interest rate, we observe clear evidence of a pronounced distributional shift across the three selected months during both the Great Recession and the COVID-19 recession. In particular, following the COVID-19 recession (December 2020), the distribution of the short-term rate becomes less symmetric and exhibits increased negative skewness.

Turning to the long-term interest rate, distributional changes appear more pronounced during the COVID-19 recession than during the Great Recession. After the Great Recession (December 2009), the distribution becomes more negatively skewed. In contrast, during the COVID-19 recession, the distribution shifts towards positive skewness, indicating a different form of market adjustment at the long end of the yield curve.

\begin{figure}[H]
\includegraphics[width=1\textwidth]{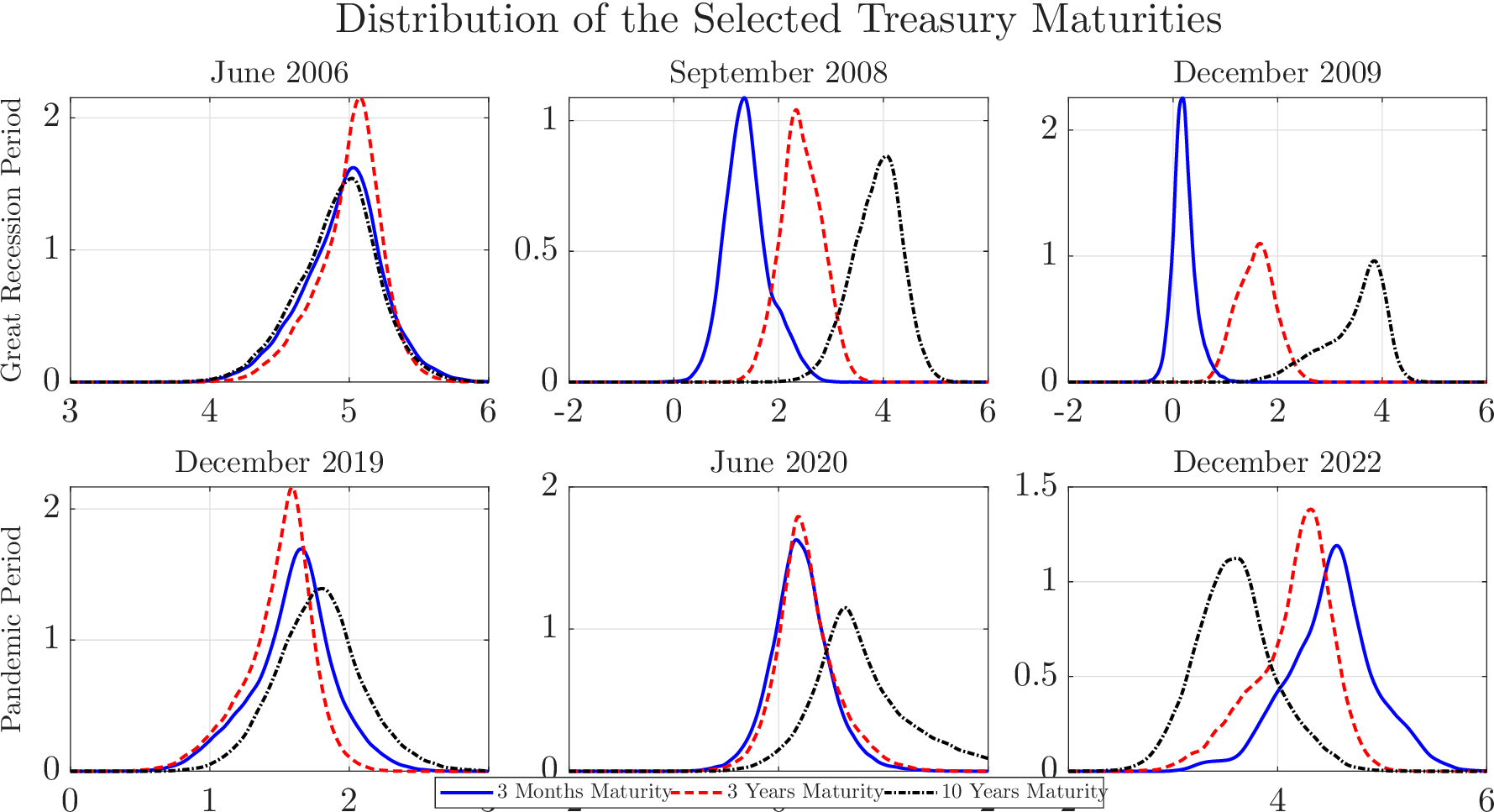}
\caption{Distribution of the 3-month, 3-year, and 10-year Treasury yields during the Great Recession (top panel) and the COVID-19 pandemic period (bottom panel). The solid blue line corresponds to the 3-month maturity, the red dashed line represents the 3-year maturity, and the black dashed line denotes the 10-year maturity.}
\label{fig:selected_density}
\end{figure}

Figure \ref{fig:selected_density} presents the estimated conditional distributions of three selected Treasury maturities—3 months, 3 years, and 10 years—representing the short, intermediate, and long segments of the yield curve, respectively, across the two crisis periods. Prior to the Great Recession, the distributions of these maturities are relatively similar. However, in the aftermath of the Great Recession, we observe a clear location shift across all maturities, with the 10-year maturity displaying increased left-skewness. During the COVID-19 recession, only the long end of the yield curve undergoes a location shift, accompanied by greater dispersion. Following the COVID-19 episode, all three maturities exhibit a location shift, with the short and intermediate maturities also displaying increased negative skewness.

These results indicate that short- and long-term interest rates, as well as selected yield maturities, experience heightened skewness and asymmetry during recessionary periods. This highlights the empirical relevance of quantile-based approaches, which allow for a more flexible representation of the conditional distribution—particularly in the tails—than traditional factor models such as Nelson-Siegel, which impose symmetry and are limited in their ability to capture such nonlinearities.

\subsubsection{The Interaction between the Yield Curve and Macroeconomy}

\begin{figure}[H]
\includegraphics[width=1\textwidth]{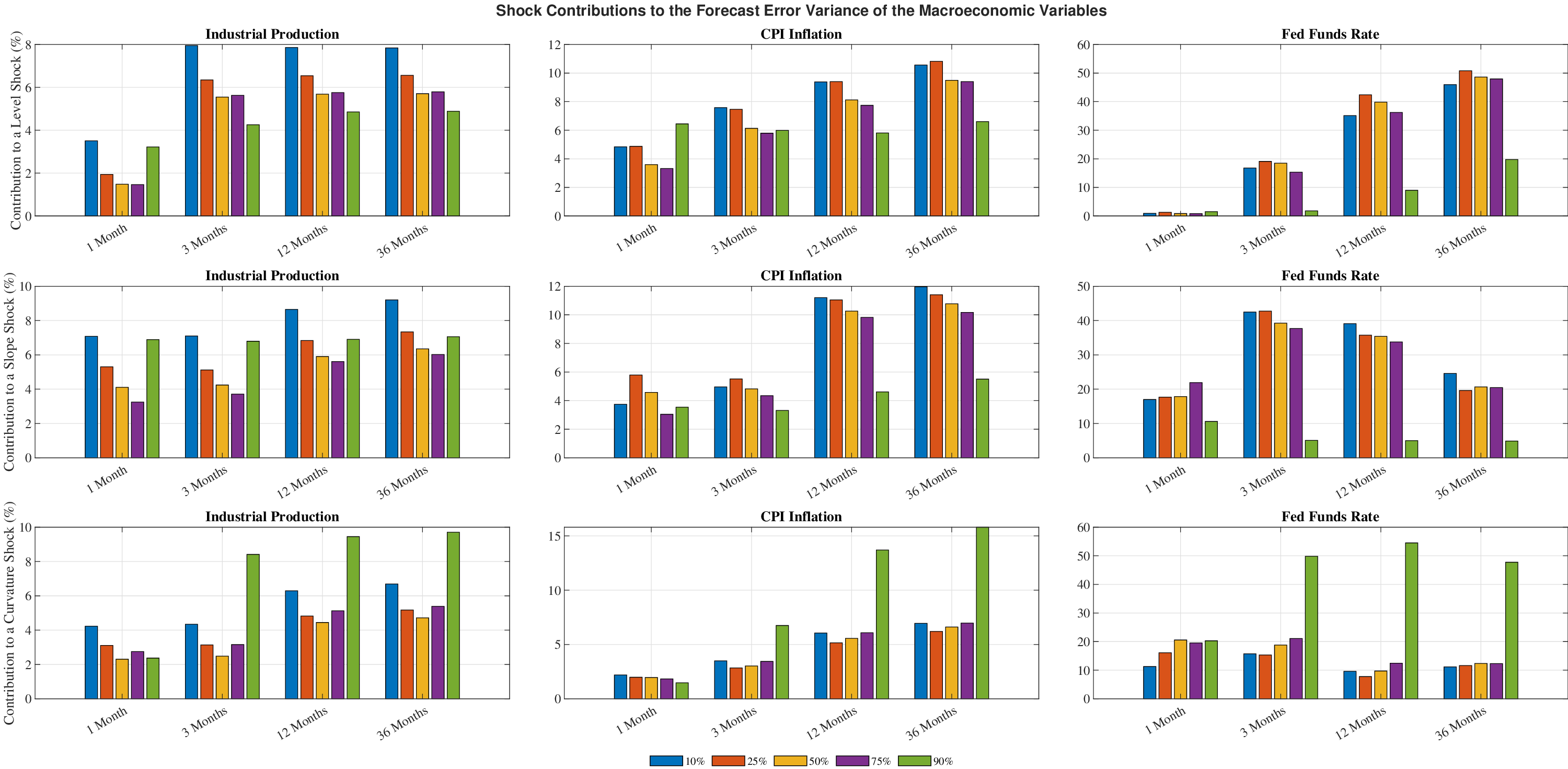}
\caption{Posterior estimates of the generalized forecast error variance decomposition for the macroeconomic variables (columns) on Yield Curve Factors (rows). The colored bars represent different quantiles: blue (10\%), orange (25\%), yellow (50\%), purple (75\%), and green (90\%).}
\label{fig:GFEVDmacro}
\end{figure}

\begin{figure}[H]
\includegraphics[width=1\textwidth]{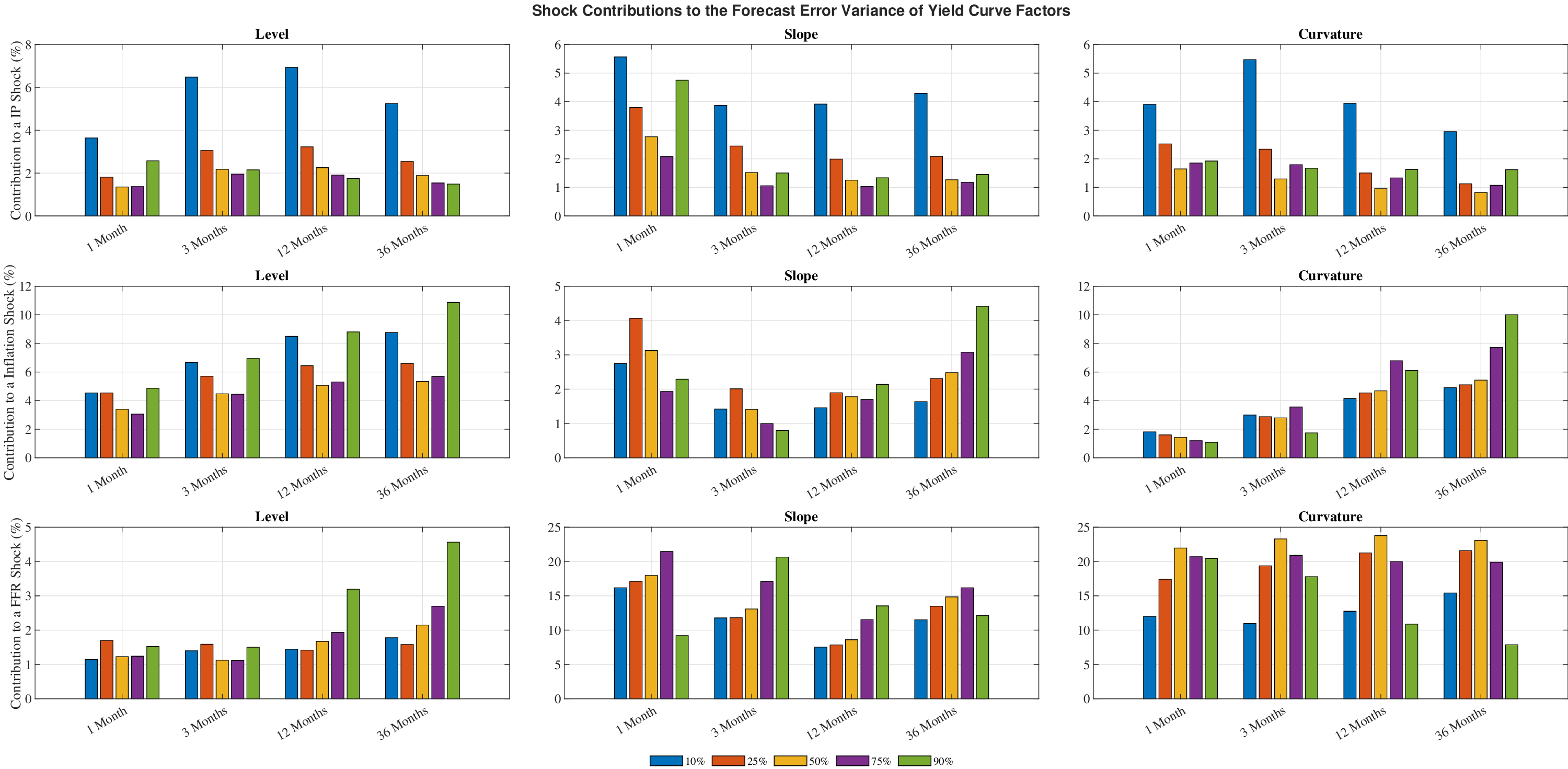}
\caption{Posterior estimates of the generalized forecast error variance decomposition for the Yield Curve Factors (columns) on Yield Curve Factors (rows). The colored bars represent different quantiles: blue (10\%), orange (25\%), yellow (50\%), purple (75\%), and green (90\%).}
\label{fig:GFEVDyield}
\end{figure}

The proposed framework, as specified in Equation~\eqref{eq:TVP_QRNS_model}, enables an in-depth analysis of the dynamic interaction between yield curve factors and macroeconomic variables by incorporating a VAR-like state equation. To explore these interactions, we compute the generalized forecast error variance decomposition (GFEVD) across five quantiles, evaluating the contribution of shocks to yield curve factors on macroeconomic variables—and vice versa—as illustrated in Figures~\ref{fig:GFEVDmacro} and~\ref{fig:GFEVDyield}, respectively.

We begin by examining the effects of yield curve shocks on macroeconomic variables (Figure~\ref{fig:GFEVDmacro}). The analysis reveals that curvature shocks generate heterogeneous impacts across the distribution of macroeconomic outcomes. Notably, these shocks have a more pronounced effect on the upper tail of all three macroeconomic variables—particularly at longer forecast horizons—indicating heightened sensitivity in more expansionary states of the economy. In contrast, level shocks increasingly influence the lower tail of industrial production over time, suggesting growing downside risks in response to shifts in the overall level of the yield curve.

We next assess the influence of macroeconomic shocks on the yield curve (Figure~\ref{fig:GFEVDyield}). Shocks to industrial production significantly affect the lower tails of all three yield curve factors, implying that negative real activity shocks disproportionately compress the yield distribution. By contrast, inflation shocks primarily impact the upper tails of the slope and curvature factors, with these effects intensifying at longer horizons—consistent with rising term premia during inflationary episodes. Additionally, federal funds rate shocks increasingly affect the upper tail of the level factor over time, underscoring the asymmetric transmission of monetary policy to the term structure.

In summary, these results provide robust evidence that the interaction between yield curve factors and macroeconomic variables is highly dependent on the quantile considered. This highlights the importance of employing quantile-based approaches, which can capture distributional asymmetries and tail behavior that standard linear models may overlook.

\section{Conclusions}  \label{sec:conclusion}

This study introduces a novel time-varying parameter quantile Nelson-Siegel (TVP-QR-NS) model, estimated within a Bayesian framework, to analyse the US term structure of interest rates. The out-of-sample forecasting results demonstrate that the TVP-QR-NS model consistently outperforms conventional benchmarks in forecasting tail risks, particularly at longer horizons and for long-term maturities. Importantly, it is the only model consistently included in the model confidence sets for these maturities and horizons, while standard conditional mean models are frequently excluded. The in-sample analysis reveals significant dispersion in yield curve factors across quantiles, especially following recessionary periods. The analysis also uncovers pronounced shifts in the distributions of short- and long-term interest rates—as well as yields at selected maturities—during the Great Recession and the COVID-19 pandemic, highlighting the importance of capturing variation across both time and quantiles. In addition, strong evidence is found of quantile-dependent dynamics in the relationship between yield curve factors and macroeconomic conditions. Overall, these findings highlight the robustness and practical relevance of the TVP-QR-NS framework for risk management and macro-financial policy analysis.

\bibliographystyle{chicago}
\bibliography{biblio.bib}

\clearpage

\appendix

\section{Appendix}

\subsection{Sampling the Decay Parameter}

As mentioned above, it has so far been common practice to fix $\lambda$ to be a particular value. On the other hand, we are agnostic and estimate $\lambda(\tau)$ given the data for each quantile. Given a uniform prior for $\lambda(\tau)$, $p(\lambda(\tau))\sim U(a,b)$, the conditional posterior of $(\lambda(\tau)|\bullet)$ is given by
\[
(\lambda(\tau)|\bullet)\propto p(\bY, \bw | \bbeta)p(\lambda(\tau)),
\]
where $a<\lambda(\tau)<b$. Since the support of this conditional density is bounded and non-standard, we draw from this conditional density using a Griddy-Gibbs step \citep{ritter1992facilitating}. This is done using the following steps:

\begin{enumerate}
	\item Construct a grid with grid points $\hat{\lambda(\tau)}_{1},\ldots,\hat{\lambda(\tau)}_{R}$,
	where $\hat{\lambda(\tau)}_{1}=a$ and $\hat{\lambda(\tau)}_{R}=b$.
	\item Compute $F_{i}=\sum_{j=1}^{i}p(\hat{\lambda(\tau)}_{j}|\bullet)$.
	\item Generate $U$ from a standard uniform distribution.
	\item Find the smallest positive integer $k$ such that $F_{k}\geqslant U$
	and return $\lambda(\tau)=\hat{\lambda(\tau)}_{k}$.
\end{enumerate}

\subsection{Additional In-Sample Results}

\begin{figure}[H]
\includegraphics[width=1\textwidth]{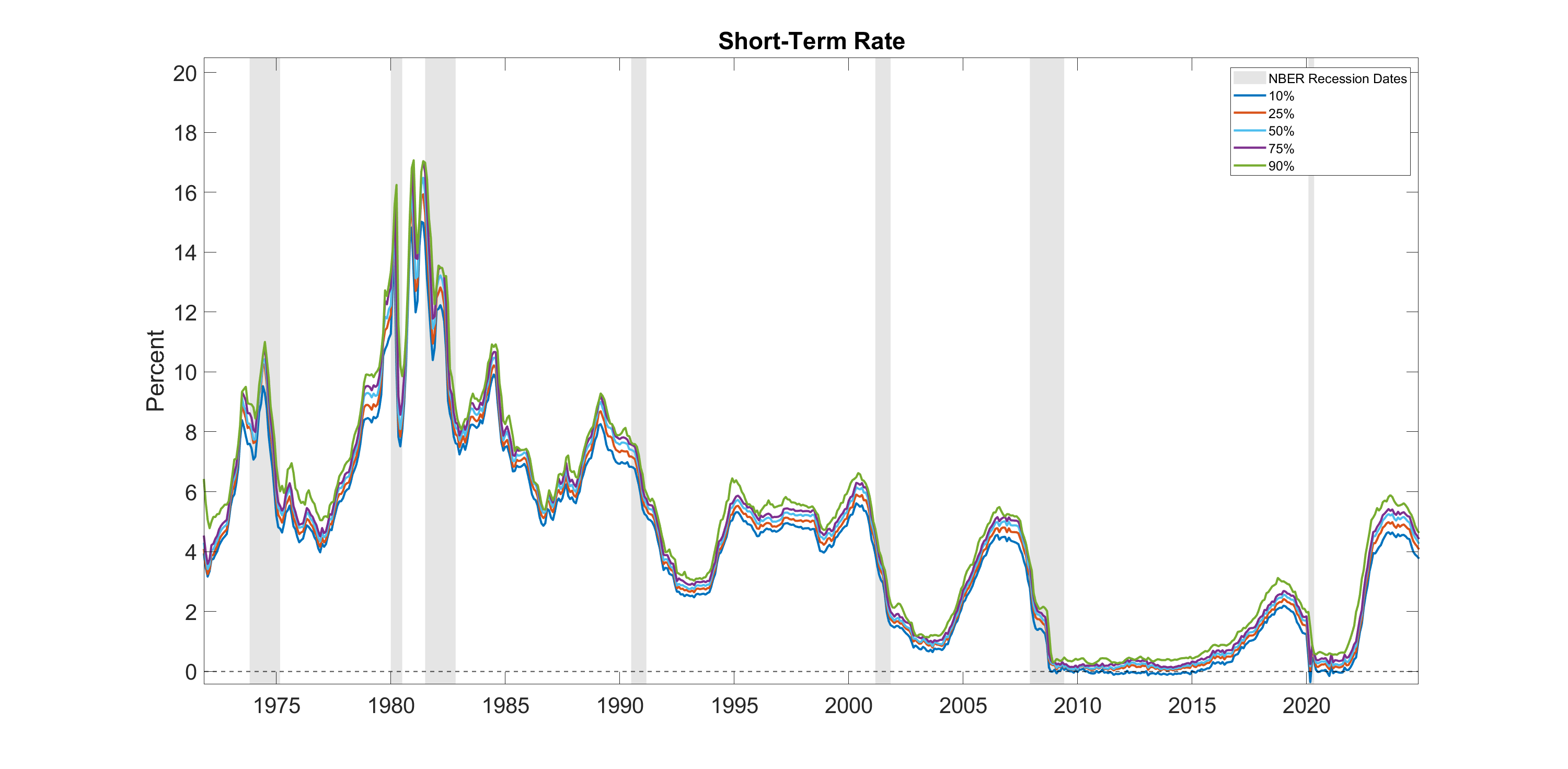}
\caption{Posterior mean estimates of the short-term rate (Level + Slope) for each of the five quantiles (10\%, 25\%, 50\%, 75\%, 90\%). NBER recession periods in grey shades.}
\label{fig:short}
\end{figure}

\end{document}